\documentclass[twocolumn,longbigliography, aps,prb,showpacs,byrevtex,amsmath,amssymb,superscriptaddress]{revtex4-1}

\usepackage{graphicx}
\usepackage{epstopdf}
\usepackage{dcolumn}
\usepackage{bm}
\usepackage{gensymb}
\usepackage{upgreek}
\usepackage{hyperref}
\usepackage{graphicx}
\usepackage{epstopdf}
\usepackage{dcolumn}
\usepackage{bm}
\usepackage{gensymb}
\usepackage{upgreek}
\usepackage{booktabs}
\usepackage{hyperref}

\usepackage{amssymb,mathtools}
\usepackage{color}
\usepackage{amsmath} 

\usepackage{tikz,xcolor,hyperref}
\definecolor{lime}{HTML}{A6CE39}
\DeclareRobustCommand{\orcidicon}{%
	\begin{tikzpicture}
	\draw[lime, fill=lime] (0,0)
	circle [radius=0.16]
	node[white] {{\fontfamily{qag}\selectfont \tiny ID}};
	\draw[white, fill=white] (-0.0625,0.095)
	circle [radius=0.007];
	\end{tikzpicture}
	\hspace{-2mm}
}

\foreach \x in {A, ..., Z}{%
	\expandafter\xdef\csname orcid\x\endcsname{\noexpand\href{https://orcid.org/\csname orcidauthor\x\endcsname}{\noexpand\orcidicon}}
}

\begin{document}

\title{Pentagonal nanowires from topological crystalline insulators:\\ a platform for intrinsic core-shell nanowires and higher-order topology}

\author{Ghulam Hussain\orcidE}
\affiliation{International Research Centre MagTop, Institute of Physics, Polish Academy of Sciences, Aleja Lotnik\'ow 32/46, PL-02668 Warsaw, Poland}
\affiliation{Institute for Advanced Study, Shenzhen University, Shenzhen 518060, China}

\author{Giuseppe Cuono\orcidD}
\affiliation{International Research Centre MagTop, Institute of Physics, Polish Academy of Sciences, Aleja Lotnik\'ow 32/46, PL-02668 Warsaw, Poland}
\affiliation{Consiglio Nazionale delle Ricerche (CNR-SPIN), Unit\'a di Ricerca presso Terzi c/o Universit\'a “G. D’Annunzio”, 66100 Chieti, Italy}

\author{Piotr Dziawa\orcidP}
\affiliation{Institute of Physics, Polish Academy of Sciences, Aleja Lotnik\'ow 32/46, PL-02668 Warsaw, Poland}

\author{Dorota Janaszko\orcidJ}
\affiliation{Institute of Physics, Polish Academy of Sciences, Aleja Lotnik\'ow 32/46, PL-02668 Warsaw, Poland}

\author{Janusz Sadowski\orcidZ}
\affiliation{Institute of Physics, Polish Academy of Sciences, Aleja Lotnik\'ow 32/46, PL-02668 Warsaw, Poland}

\author{Slawomir Kret\orcidS}
\affiliation{Institute of Physics, Polish Academy of Sciences, Aleja Lotnik\'ow 32/46, PL-02668 Warsaw, Poland}

\author{Bogus{\l}awa Kurowska\orcidB}
\affiliation{Institute of Physics, Polish Academy of Sciences, Aleja Lotnik\'ow 32/46, PL-02668 Warsaw, Poland}

\author{Jakub Polaczy\'nski\orcidK} 
\affiliation{International Research Centre MagTop, Institute of Physics, Polish Academy of Sciences, Aleja Lotnik\'ow 32/46, PL-02668 Warsaw, Poland}

\author{Kinga Warda}
\affiliation{International Research Centre MagTop, Institute of Physics, Polish Academy of Sciences, Aleja Lotnik\'ow 32/46, PL-02668 Warsaw, Poland}
\affiliation{Faculty of Applied Physics and Mathematics, Gdansk University of Technology, Gda\'nsk 80-233, Poland}

\author{Shahid Sattar}
\affiliation{Department of Physics and Electrical Engineering, Linnaeus University, 392 31 Kalmar, Sweden}

\author{Carlo M. Canali}
\affiliation{Department of Physics and Electrical Engineering, Linnaeus University, 392 31 Kalmar, Sweden}

\author{Alexander Lau}
\affiliation{International Research Centre MagTop, Institute of Physics, Polish Academy of Sciences, Aleja Lotnik\'ow 32/46, PL-02668 Warsaw, Poland}

\author{Wojciech Brzezicki\orcidW}
\affiliation{International Research Centre MagTop, Institute of Physics, Polish Academy of Sciences, Aleja Lotnik\'ow 32/46, PL-02668 Warsaw, Poland}

\author{Tomasz Story\orcidT}
\affiliation{International Research Centre MagTop, Institute of Physics, Polish Academy of Sciences, Aleja Lotnik\'ow 32/46, PL-02668 Warsaw, Poland}
\affiliation{Institute of Physics, Polish Academy of Sciences, Aleja Lotnik\'ow 32/46, PL-02668 Warsaw, Poland}

\author{Carmine Autieri\orcidA}
\email{autieri@magtop.ifpan.edu.pl}
\affiliation{International Research Centre MagTop, Institute of Physics, Polish Academy of Sciences, Aleja Lotnik\'ow 32/46, PL-02668 Warsaw, Poland}

\begin{abstract}We report on the experimental realization of  Pb$_{1-x}$Sn$_x$Te pentagonal nanowires (NWs) with [110] orientation using molecular beam epitaxy techniques. Using first-principles calculations, we investigate the structural stability in NWs of SnTe and PbTe in three different structural phases: cubic, pentagonal with [001] orientation and pentagonal with [110] orientation. Within a semiclassical approach, we show that the interplay between ionic and covalent bonds favors the formation of pentagonal NWs. Additionally, we find that this pentagonal structure is more likely to occur in tellurides than in selenides. The disclination and twin boundary cause the electronic states originating from the NW core region to generate a conducting band connecting the valence and conduction bands, creating a symmetry-enforced metallic phase. The metallic core band has opposite slopes in the cases of Sn and Te twin boundary, while the bands from the shell are insulating. 
We finally study the electronic and topological properties of pentagonal NWs unveiling their potential as a new platform for higher-order topology and fractional charge.
These pentagonal NWs represent a unique case of intrinsic core-shell one-dimensional nanostructures with distinct structural, electronic and topological properties between the core and the shell region.\end{abstract}

\date{\today}
\maketitle

\section{Introduction}
Three-dimensional topological insulators have gapped bulk states but gapless conducting edge states dictated by the bulk topological invariant and protected by the relevant symmetries. Higher-order band topological insulators (HOTIs) provide lower-dimensional boundary states as gapless hinge or corner states.\cite{Xie2021,PhysRevB.98.081110,Kooi20}
Within the magnetic topological insulators, the axion insulating phase is an example of higher-order topology proposed in compounds and superlattices\cite{PhysRevLett.122.256402,PhysRevB.107.125102,PhysRevB.103.195308}.
Very few real materials have been proposed for the experimental realization of HOTI, two of the most popular are bismuth\cite{Schindler18Naturephysics} and cubic SnTe nanowires (NWs)\cite{Schindler18}. Indeed, it was proposed that the cubic SnTe NWs under rhombohedral strain can host a helical higher-order topological insulator phase \cite{Schindler18}. This is protected by both mirror and time-reversal symmetry, but no inversion symmetry is needed for helical HOTI in SnTe. The rhombohedral strain has the role of suppressing the first-order topology to make the second-order topology visible\cite{Schindler18}. Recently, it was proposed a new route through defects or dislocations to obtain higher-order topological insulators\cite{10.21468/SciPostPhys.10.4.092,Schindler22,PhysRevB.89.224503,PhysRevResearch.3.033107} or fractional charges\cite{Peterson2021,Liu2021}. In case of disclinations, the system can host
strong second-order topological phases provided that local C$_2$, C$_4$, or C$_6$ symmetries protecting the topological phase are present\cite{10.21468/SciPostPhys.10.4.092}; the same happens for one-dimensional disclinations\cite{PhysRevB.107.214108}.\\

Group IV-VI semiconductors are known to display interesting characteristics, including thermoelectric effects \cite{wang2011heavily, wood1988materials}, ferroelectric behavior \cite{lebedev1994ferroelectric, liu2020synthesis}, and superconducting properties \cite{matsushita2006type, mazur2019experimental}.
Notably, the identification of a topological crystalline insulating phase (TCI) in materials such as SnTe \cite{fu2011topological, hsieh2012topological,Lau19,Tanaka07,Okada13}, along with certain related substitutional compounds such as Pb$_{1-x}$Sn$_x$Te \cite{xu2012observation} and Pb$_{1-x}$Sn$_x$Se \cite{dziawa2012topological,Lusakowski23}, has sparked significant interest in the realm of research. 
This category of materials has garnered substantial attention for further investigation. The TCI phase stands apart from the conventional topological insulating phase, as the linearly dispersing Dirac states on the high-symmetry surfaces of TCIs are protected by crystal symmetries\cite{Fu07} rather than time-reversal symmetry. Consequently, the topological invariant is also different. For example, in the case of SnTe and its alloys, the invariant is named mirror Chern number \cite{Teo08,Safaei_2015,hsieh2012topological,Cuono22,Rechcinski21} since the symmetry which protects the surface states is a reflection of the crystal with respect to (110) mirror plane. 
Changing the dimensions or the size (whether in a 2D or 3D configuration) of materials results in the modification of their distinctive properties. Notably, a transition from bulk \cite{Plekhanov14,Wang20,Barone13} to thin films \cite{Slawinska20,Volobuev17,Liu18} induces property variations. A case in point is SnTe: while it behaves as a trivial insulator at reduced thicknesses, it assumes a topological nature beyond a critical thickness\cite{liu2014spin,liu2015crystal,PhysRevB.90.045309,Safaei_2015,PhysRevB.91.081407}, displaying strong robustness against impurity doping\cite{D1NR07120C}.
Employing scanning tunneling microscopy and spectroscopy on Pb$_{1-x}$Sn$_x$Se, it was demonstrated that the (001) surface hosts large terraces interspersed with step edges of varying heights. The features of low-energy states occurring at surface atomic steps have been investigated, along with an exploration of the response of these edge channels to the stoichiometry of the alloy\cite{Sessi16,Brzezicki19,mazur2019experimental,Wagner12,PhysRevB.98.245302}.

\begin{figure*}[t]	
\centering
\includegraphics[width=2\columnwidth, angle=0]{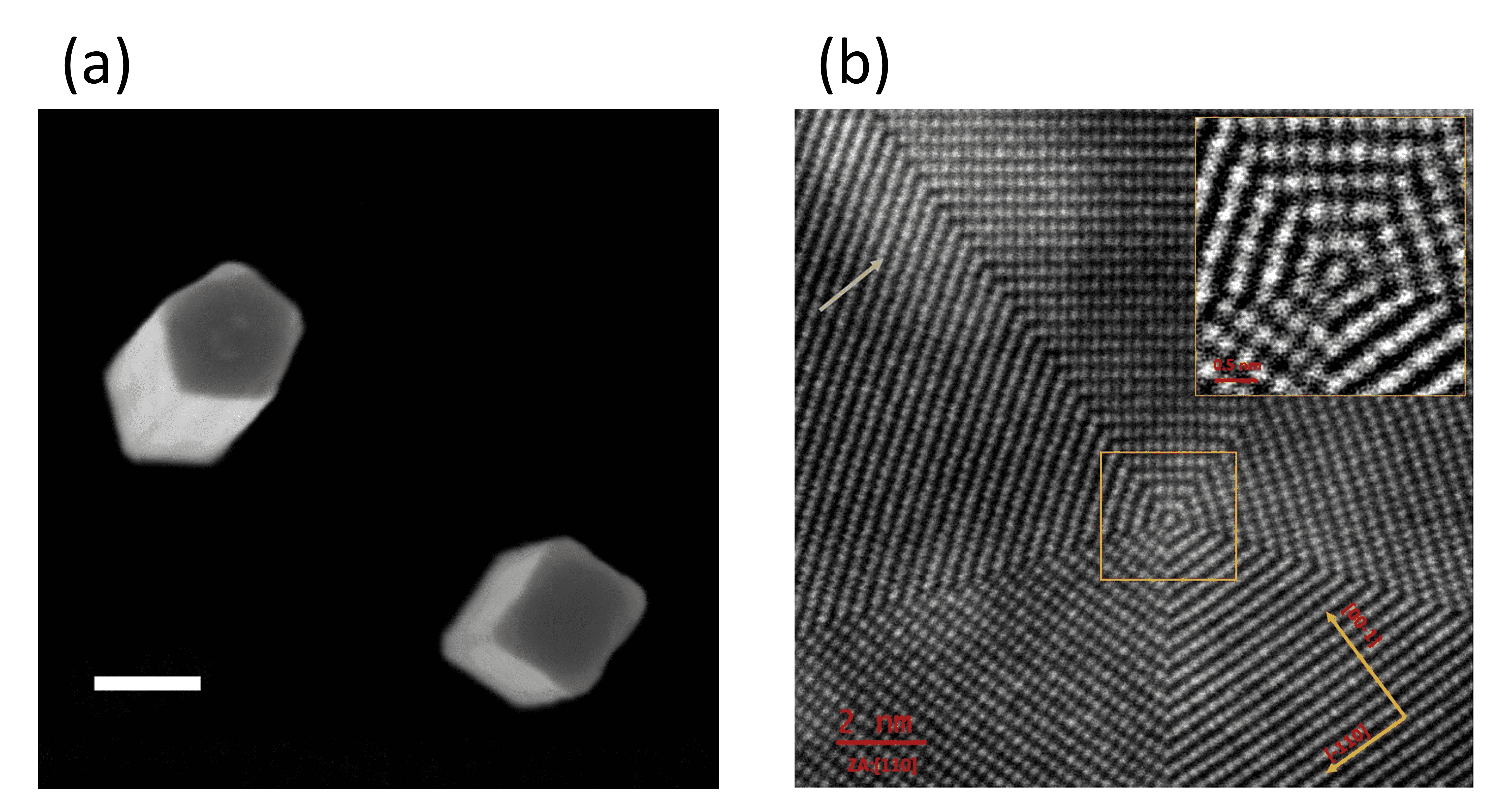}
\caption{(a) Scanning electron microscopy (SEM) top view image of Pb$_{1-x}$Sn$_{x}$Te NWs at magnification 300k presenting the two possible types of cross-sections. The white scale bar is 100 nm. (b) STEM image of five-fold Pb$_{1-x}$Sn$_{x}$Te NW with visible cubic rock salt crystal structure, and orange marked the center of the structure is zoomed in the right corner as an inset.  The inset in the right corner shows a disclination with a Frank angle of +90$^\circ$ and a core chain in the center of the pentagon. The arrow indicates brighter areas related to strain fields on the twin boundary.}\label{Experimental}
\end{figure*} 

The impacts resulting from the existence of defects in the TCIs were also investigated in the literature. 
Possible defects are disclinations or dislocations, which naturally occur in the material during its growth. These are defects that interrupt the rotational or translational symmetry in a region of the material. Fractional electric charges were observed in disclinations \cite{Ortix16Nature,Peterson2021,Liu21Nature}, it was shown that these defects can probe higher-order topology \cite{Schindler18,Schindler18Naturephysics,Roy21}, and the criteria for a zero-energy Majorana bound state to be present at defects was given \cite{Benalcazar14,Chen22}. Partial-dislocation-induced topological modes in 2D and 3D higher-order topological insulators were experimentally observed \cite{Yamada22}.
In the recent past, an investigation was undertaken to explore the connection between the twinning in thin films and variations in the mirror Chern number\cite{Samadi23}.

Exploring the one-dimensional (1D) case,  the topological crystalline insulator SnTe NWs emerge as even more captivating than their bulk counterparts. This enhanced intrigue arises from the amplified contribution originating from topological surface states and the effects of confinement in 1D systems\cite{C4NR05124F,C4NR05870D,safdar2013topological,Li2013single}.
Diverse approaches were uncovered to produce SnTe NWs. These methods encompass the utilization of high-yield alloy nanoparticles as growth catalysts \cite{liu2020synthesis}, graphene \cite{Sadowski2018} and employing single-walled carbon nanotubes to produce the smallest possible SnTe NWs \cite{Vasylenko18}.
SnTe NWs have also been thoroughly investigated theoretically. It was shown that it is possible to obtain distinct topological states through the application of varied fields \cite{Nguyen22}. Notably, bulk Majorana modes appear in the presence of inversion symmetry, while the imposition of symmetry-breaking fields at the NW ends leads to the emergence of Majorana zero modes, alongside the gapping of the Majorana bulk modes \cite{Nguyen22}. 
Cubic Pb$_{1-x}$Sn$_{x}$Te NWs have also been synthesized \cite{Safdar15,dad2022nearly,Saghir15,Xu2016-vw}. TCI NWs can serve as a platform to confine and control Dirac fermions \cite{Skiff22}, while PbTe NWs have drawn attention due to their potential to host Majorana zero modes \cite{Song23,Jung22}.
However, the realization of Majorana’s fermions is challenging as it requires
the growth of NWs to be integrated with a circuit that exhibits the necessary sensitivity to monitor individual electrons as they traverse through it. All of this intricate work must be conducted at extremely low temperatures and within a magnetic field that is 10,000 times stronger than Earth’s field\cite{Frolov21}.

{\it Ab initio} calculations were applied to investigate ultrathin cubic NWs\cite{Shukla22,Hussain23cubic} and to find the critical thicknesses at which SnTe NWs go from a trivial insulating regime to a spin-orbit insulating phase and then to the TCI phase \cite{Hussain23cubic}.
Great attention was recently dedicated to 2D pentagonal materials in different types of the Cairo pentagonal tessellation\cite{SHEN20221,WINCZEWSKI2022110914,doi:10.1021/acsnano.1c04325} and their topological properties\cite{Bravo2019}. 
Studies are present also for 1D pentagonal materials referred as penta-twinned nanowires; however, only elemental and non-ionic pentagonal NWs have been experimentally realized\cite{PhysRevLett.93.126103,REYESGASGA2006162,ANTSOV2019102686} or theoretically proposed\cite{Sen02,Ma13,Sainath16,Neon_NW}. 

In this paper, we study the structural, electronic and topological properties of the SnTe pentagonal NWs with [110] orientation. 
Respect to the literature\cite{Peterson2021}, our system differs for the [110] orientation, for the periodicity along the z-direction and for hosting a core chain (CC) that produces peculiar properties.
First, we report the experimental realization of Pb$_{1-x}$Sn$_x$Te ionic pentagonal NWs using molecular beam epitaxy technique. Then we investigate, by means of {\it ab initio} and model Hamiltonian calculations, the structural, electronic and topological properties of pentagonal PbTe and SnTe NWs with anionic and cationic twin boundaries.
We find that the cubic [001] NWs show a robust insulating behavior \cite{Hussain23cubic}, while the pentagonal NWs are metastable with a more covalent bond. This more covalent bond brings smaller band gaps and therefore favors the topological phases.  
We investigate the evolution of the band gap as a function of the thickness and we study the topology of the pentagonal NWs.\\
The paper is organized as follows: in the next Section the experimental realization of the ionic pentagonal NWs is described, and the third section is instead devoted to the theoretical investigation of the structural, electronic and topological properties of the pentagonal NWs. In the fourth and last Section, we draw our conclusions.

\section{Experimental realization of ionic pentagonal nanowires}

Pb$_{1-x}$Sn$_{x}$Te NWs were grown using the molecular beam epitaxy (MBE) approach within a home-built apparatus furnished with solid sources of SnTe, Pb, and Te (refer to Appendix A in the Supp. Information for more comprehensive information). The NWs obtained with this method have typical lengths exceeding 1 $\mu$m, with a NW growth rate surpassing 0.5 $\mu$m/h. The NW diameters display a relatively broad range, spanning from a minimum of approximately 15 nm to as much as 200 nm. The peak distribution rests at around 40-80 nm. Under these specific conditions, both varieties of NWs are observed: those manifesting a four-fold and those with a five-fold symmetry in their cross-sectional profiles (see Fig. \ref{Experimental}a). Utilizing a Scanning Electron Microscope (SEM), we conducted observations at 5 kV, with magnifications ranging from 10k to 300k x. The SEM imagery reveals a diverse NW length, spanning from 0.7 $\mu$m to 3 $\mu$m, while the width of the structures spans between 40 nm and 150 nm. On the substrate, a prevalent occurrence of five-fold NWs was noted, with less frequent instances of four-fold NWs. These structures exhibit remarkable symmetry and well-defined shapes, displaying evident blocks. The NW surfaces exhibit a flawlessly smooth texture devoid of imperfections.

To explore the structural attributes of the Pb$_{1-x}$Sn$_{x}$Te NWs, we employed a transmission electron microscope. This analysis involved cross-sectioned NWs inspected along the [110] zone axis, which aligns with the NW growth axis. Both scanning transmission electron microscopy (STEM) and high-resolution transmission electron microscopy modes were employed. The presence of five-fold symmetry was conclusively established. Furthermore, distinct features such as the (111) plane twin boundaries were discernible, along with brighter areas indicative of twinning. These areas are related to strain fields. In Fig. \ref{Experimental}b), we show the STEM top view image of the five-fold Pb$_{1-x}$Sn$_{x}$Te NW. At the core of the pentagonal structure, we have a one-dimensional disclination that follows the length of the pentagonal NW. The disclination is described by the Frank angle\cite{Peterson2021}, which in this case is +90$^\circ$. Despite having cubic domains between the twin boundaries, the pentagonal NWs present a 1D disclination and five twin-boundary planes; even if these planes are extended defects, their effect on the structure is local so that most of the whole volume of the NW is in cubic arrangement making this phase equivalent to the cubic structure from the thermodynamic point of view\cite{Marks_2016}. From the experimental data, the twin boundaries of the experimentally obtained pentagonal nanowires seem to be composed of anion atoms.

\begin{figure}[th]
\centering
\includegraphics[scale=0.33]{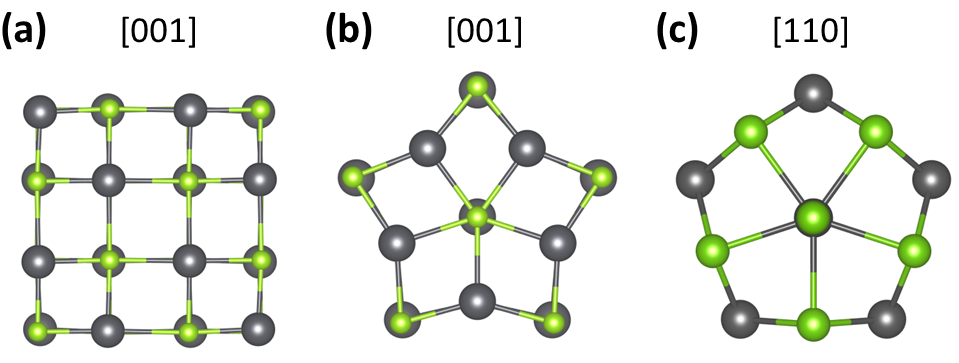}
\caption{Top view of the optimized SnTe NWs with (a) cubic structure oriented along [001] (b) pentagonal structure along [001] and (c) pentagonal structure along [110] directions. The dark grey and green balls represent the Sn and Te atoms, respectively.}
\label{Crystal structure}
\end{figure}

\begin{figure}[th]
\centering
\includegraphics[width=\columnwidth]{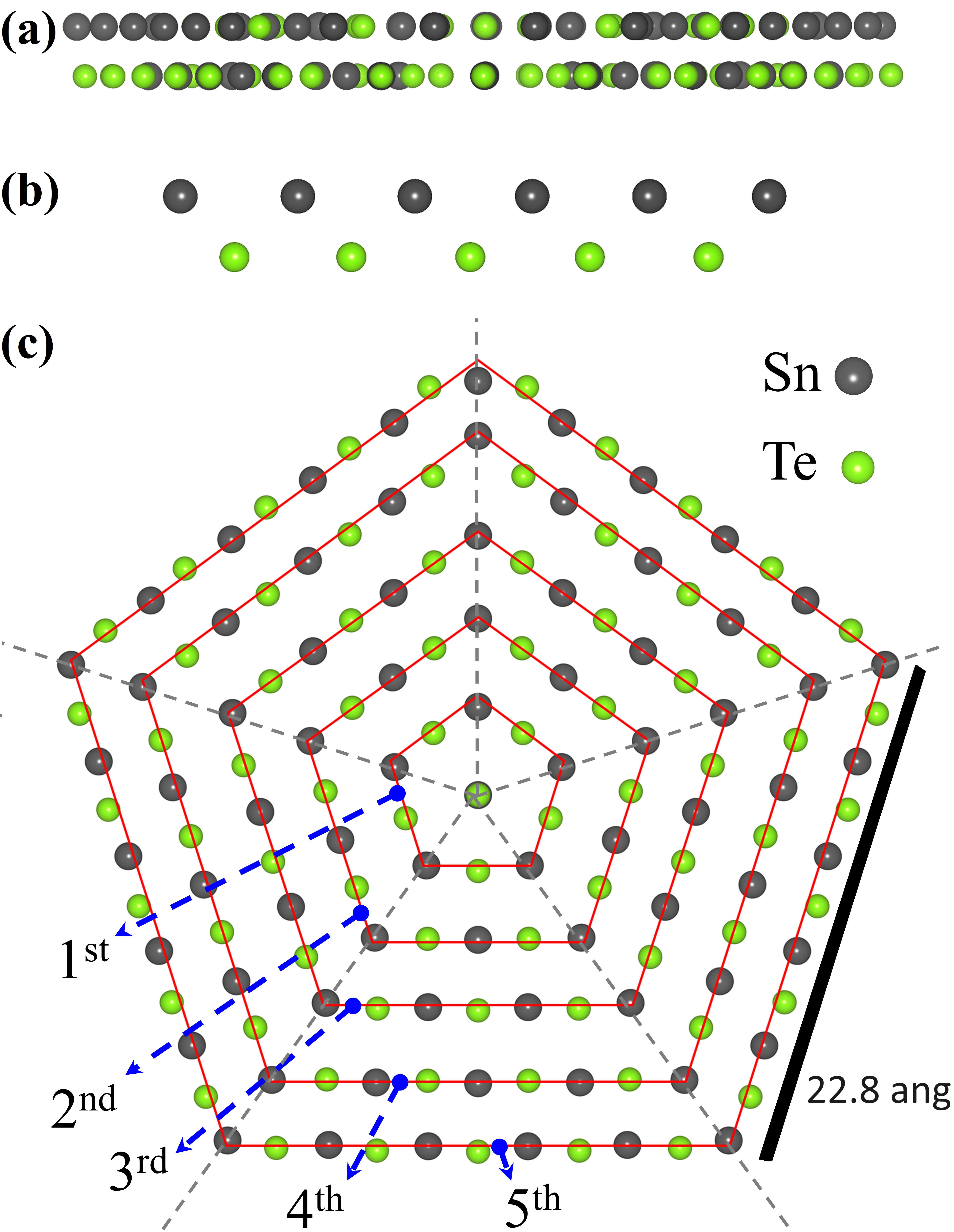}
\caption{(a-b) Side-tilted and top (c) views of SnTe pentagonal NWs oriented along [110] direction with cation twin boundary after structural relaxation. The rings are indicated with red lines, while the twin boundaries are highlighted with a dotted grey line. At the core, there is the one-dimensional disclination that lives in the entire length of the NWs. These crystal structures are stoichiometric since they host the same number of anions and cations. For the 5-ring structure of SnTe, the side of the pentagon (shown in black in the figure) is 22.8 {\AA} long. The free surfaces of the pentagonal NWs are made of (110) planes as (110) cubic NWs.}
\label{crystal_pentagonal}
\end{figure}

\begin{figure}[th]
\centering
\includegraphics[scale=0.33]{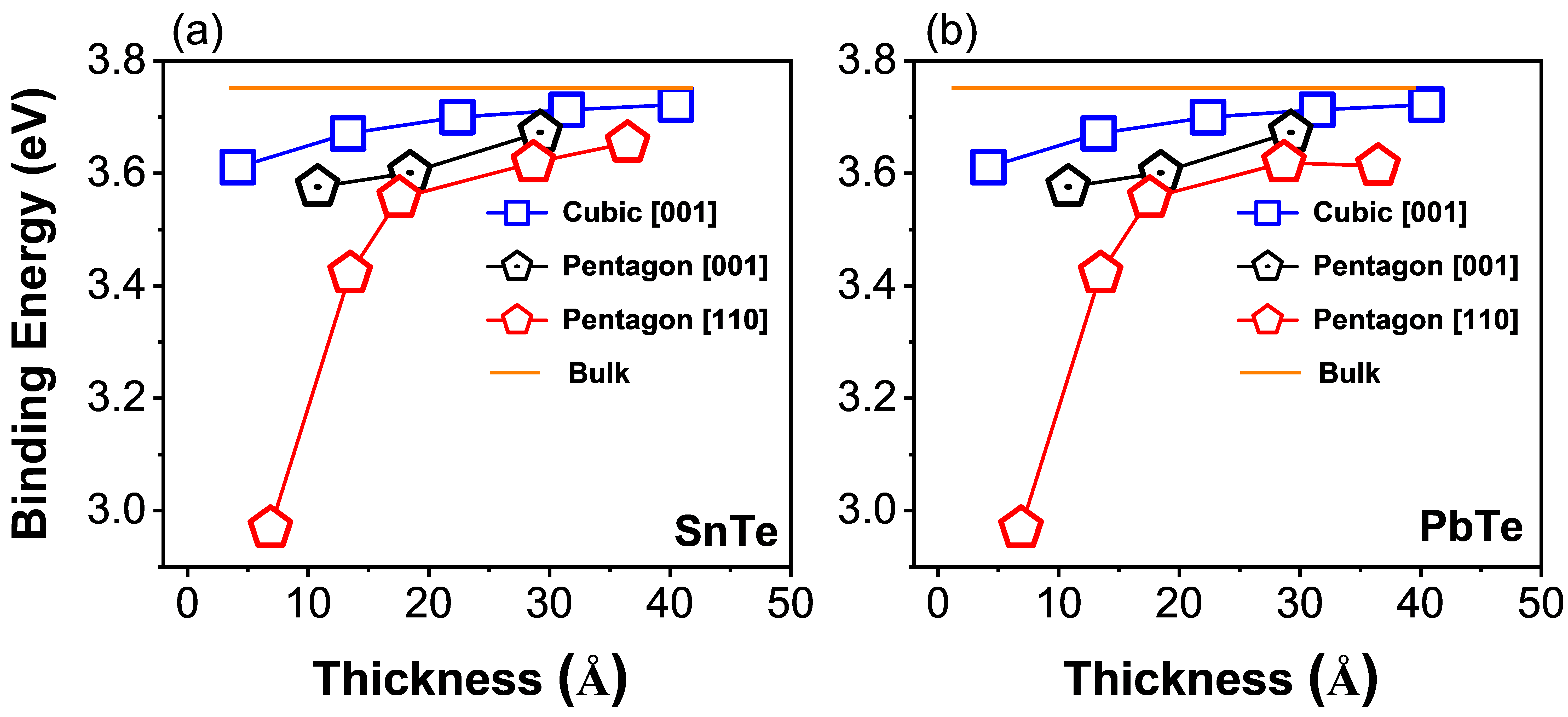}
\caption{Binding energy of (a) SnTe and (b) PbTe systems as a function of thickness for the two growth directions ([001] and [110]) without SOC. The horizontal orange line indicates the binding energy of the bulk case, while the blue cube, black and red pentagon represent the binding energy for the NWs cases. A thickness-dependent binding energy and hence stability can be observed. Cubic NWs have the highest binding energy after bulk, while pentagonal NWs bear the second-highest values.}
\label{Binding energy}
\end{figure}

\section{Structural and electronic properties}

We divide this Section into three subsections devoting the first to the study of the structural properties, the second to the electronic band structure results and the third to the topological properties.
We will work on SnTe and PbTe in the first two subsections, showing that they have the same properties at least in the thin limit. In the third and fourth sections, we will then investigate the topological properties focusing only on the SnTe NWs.

\subsection{Structural stability of the nanowires}

Using density functional theory (DFT) calculations, we compare the energetic stability of three NW crystal structures. The computational details of the calculations are reported in Appendix B in the Supp. Information.
We study the cubic NWs with [001] growth orientation and the two growth orientations [001] and [110] for the pentagonal NWs. The pentagonal structures have 5 mirror planes that coincide with the plane of the twin boundaries. The pentagonal structure with [110] orientation is the phase observed in experiments. The three optimized structures of thin SnTe NWs are illustrated in Fig. \ref{Crystal structure}. In Fig. \ref{crystal_pentagonal}, we report the side and top view of the [110] pentagonal NWs. 
The [110] pentagonal NW is composed of a core region and a shell region. The core region is composed of a core chain (CC) along the z-axis where the atoms see an environment with rotational symmetry C$_5$. 
Additionally, there is a mirror plane at z=0 or z=0.5, but there is no inversion point. 
The forces acting on the atoms of the core chain (one Sn and one Te atom per unit cell) are zero by symmetry. Indeed, the forces along the z-axis are compensated by the mirror symmetry while for the in-plane components, the sum of the forces of equivalent atoms located at the vertex of a regular pentagon on the center is zero.
The shell region is composed of the 5 twin boundaries that separate regions where the system is locally cubic and where the atoms see locally the C$_4$ symmetry present in the bulk system.
As we can see, the thickness of these NWs can be increased by adding more rings of corresponding atoms to the edges (sides) of each NW. In this way, the pentagonal NWs are named as 1-ring, 2-ring, 3-ring, 4-ring and 5-ring, respectively.

To examine the stability of such NW systems, the ground state binding energies per atom are calculated from the ground state total energies obtained via DFT\cite{Sen02,adhikari2012stabilities}:
\begin{equation*}
   E_B =\frac{\substack{n_1E_X + n_2E_Y - (E_{XY})}}{\substack{n_1 + n_2}}
\end{equation*}
Here, $n_1$ and $n_2$ are the numbers of cation atoms (Sn, Pb) and Te atoms in the NW, respectively. E$_X$ and E$_Y$ represent the ground state total energies for individual atoms (E$_{Sn}$, E$_{Pb}$, E$_{Te}$), and the term E$_{XY}$ represents the total energy of the relaxed NW i.e. E$_{SnTe}$ or E$_{PbTe}$. The higher the binding energy, the higher the structural stability. Fig. \ref{Binding energy} shows the binding energy (E$_B$) for SnTe and PbTe NWs oriented along [001] and [110] directions, as a function of NW thickness. Also, E$_B$ of the bulk system is portrayed as a line for comparison. It can be seen that E$_B$ varies with the number of atoms in the NW. As the thickness of the NW increases, the binding energy also increases. This monotonic increase of E$_B$ reflects enhancement in the stability of NWs. From Fig. \ref{Binding energy}, we observed that pentagonal NWs grown in [001] direction bear more stability and are closer to bulk value than NWs grown along [110] orientation. However, the [001] cubic structure of NWs is observed to be the most stable one as compared to other cases. Following this analysis, the cubic structure is still the ground state for SnTe and PbTe NWs as in the bulk phase, while the pentagonal structure of SnTe and PbTe NWs is a metastable state. We calculate the binding energy as a function of the number of atoms in the unit cell, and in the limit of infinite atoms, the binding energy should tend to the value of the bulk. Indeed, the binding energy differences strongly decrease as a function of the thicknesses of NWs and could reduce to a few tens of meV for thick NWs. Therefore, it is unlikely to have thin pentagonal NWs, while they are more likely to occur for large thicknesses. 
Recently, it was shown that several structural configurations are possible in the case of elemental NWs with a small difference in the binding energy of few meV\cite{TARKOWSKI2023107241}. In the ionic NWs as SnTe, the number of configurations would be certainly smaller with respect to elemental NWs due to the constraints on the alternation of anion and cation; however, as we demonstrated experimentally, it is possible to stabilize structural configurations different from the lowest energy crystal structure. Experimental information supporting the metastable property of the pentagonal NW comes from the growth temperature. As mentioned in the previous Section, the pentagonal NWs are obtained only in high-temperature experiments, where metastable phases are more likely to be obtained.

DFT calculations of the 
pentagonal nanowires with Pb$_{0.50}$Sn$_{0.50}$Te composition do not change the scenario from the point of view of the total energy, indicating that the formation of pentagonal NWs comes from a less likelihood to obtain Pb$_{0.50}$Sn$_{0.50}$Te cubic NWs. In order to further understand the stability of the pentagonal nanowires, we discuss what is peculiar in the Pb$_{0.50}$Sn$_{0.50}$Te. Well, close to this composition we have the closure of the band gap with the topological transition in the bulk samples or thick cubic NWs at room temperature\cite{Dornhaus1983}. 
The cubic structures of Pb$_{1-x}$Sn$_{x}$Te are stabilized by the ionic insulator character described by the Madelung constant, while metallic systems tend to have a more hexagonal structure to produce a large atomic packing factor and covalent bonds. 
We propose that the thick cubic Pb$_{0.50}$Sn$_{0.50}$Te NWs with zero gap are structurally less stable than SnTe and PbTe NWs, such that the pentagonal Pb$_{0.50}$Sn$_{0.50}$Te NWs are more likely to grow. 
To test this claim at the quantum mechanical level, we would need to reproduce the energetic stability of the alloyed cubic and pentagonal NWs within DFT, however, 
we are not able to simulate cubic NWs with spin-orbit above 15 nm thickness\cite{Hussain23cubic}. Therefore, we move to a semiclassical approach to provide insight on the formation of the disclination in the initial stages of the growth.

A semiclassical approach to describing the stability of ionic pentagonal NWs is reported in Appendix C in the Supp. Information. Within the semiclassical approach, for the cube, the Madelung constant is highest and particle density is lowest, indicating that the contribution of the ionic bond is larger than the covalent one. In the case of hexagon, particle density is the highest and Madelung constant is the lowest indicating that the covalent bond is the dominant character of the binding. However, the pentagonal structure bears intermediate values, which implies that if the bond is partially ionic and partially covalent, the pentagonal system is expected to be the most stable one. Basically, the pentagonal structure arises from the interplay between the ionic bond and the covalent bond. The pentagonal NW tends to be stabilized in a more covalent environment which is favored by small gaps and metallicity as happens close to the topological transition in Pb$_{0.50}$Sn$_{0.50}$Te. In the Supporting Information, we conclude that NWs based on Pb and Te would be more likely to form this unusual pentagonal structure since they will have weaker ionic bonds.

\begin{figure}[th]
\centering
\includegraphics[scale=0.05]{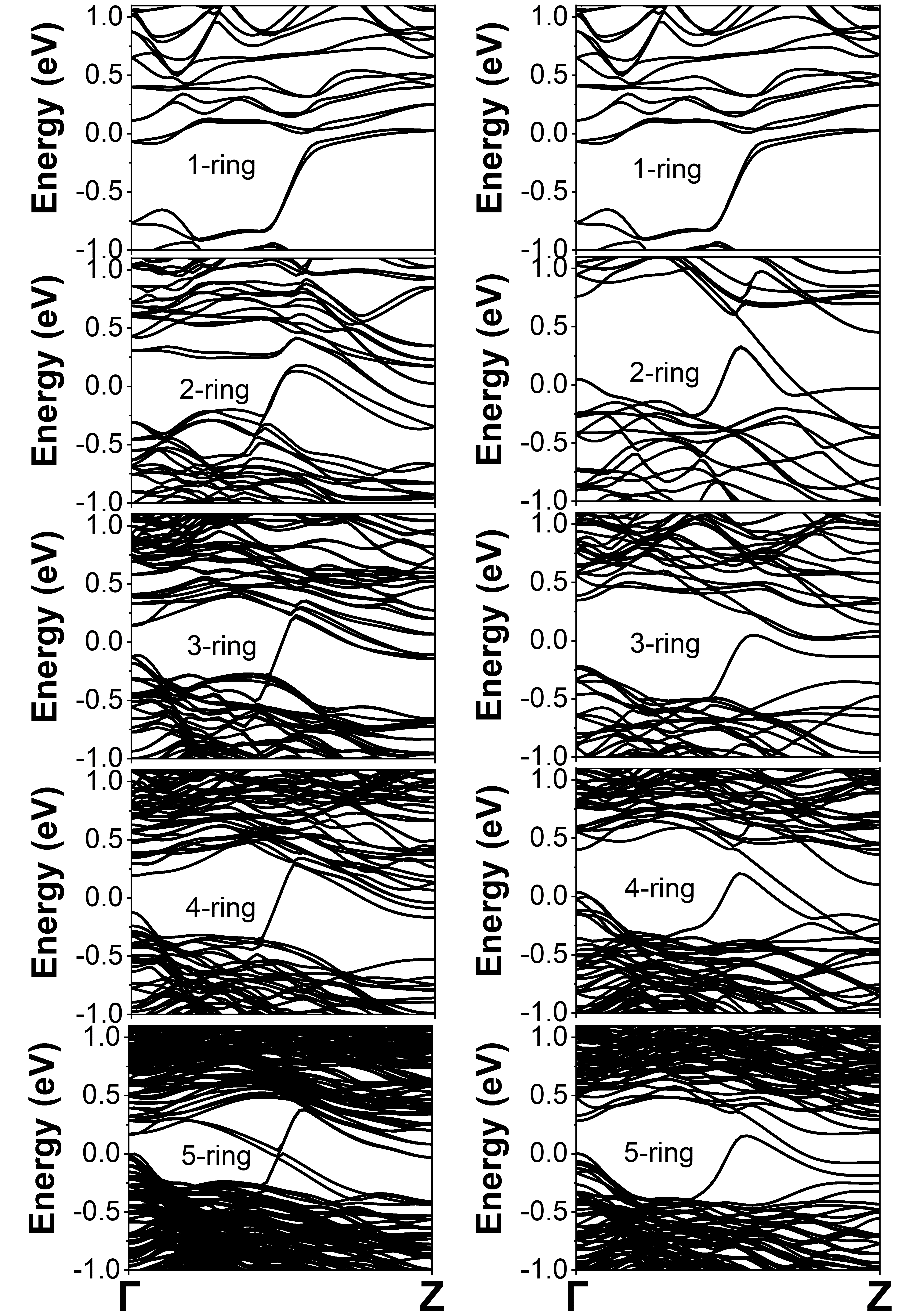}
\caption{Electronic band structures for [110] oriented pentagonal SnTe NWs with SOC included. (Left panel) Band structures with Sn on the twin boundaries, (right panel) and the results for Te on the twin boundaries. The band structures of 1-ring are equal in both cases since the structural relaxation brings to the same crystal structure.}
\label{Bands_110_SOC}
\end{figure}

Further research is necessary to advance towards the alloyed phase Pb$_{1-x}$Sn$_x$Te. Our preliminary results indicate that the difference between the binding energy of the cubic and pentagonal NWs decreases to smaller values in the alloy.
Exploring the possibility that the CC consists solely of either Sn or Pb atoms could be a direction worth investigating. In this regard, employing genetic algorithms might predict the crystal structure and the cationic composition of the CC\cite{TARKOWSKI2023107241}.

\subsection{Electronic properties of the pentagonal nanowires}

In general, as we decrease the thickness, we can observe a bandwidth reduction. The bandwidth reduction brings metals and topological systems into the trivial insulating phase\cite{vanthiel2020coupling,Autieri_2016,Hussain23cubic}. For the cubic PbTe NWs, the band gap decreases as the thickness increases.
For the ultrathin cubic SnTe NWs, we have a trivial insulating behavior with the band gap decreasing as we increase the thickness up to the appearance of a trivial spin-orbit insulating phase, and further increasing the thickness the topological phase emerges\cite{Hussain23cubic}.

To see the effect of the thickness dependence on the electronic properties of pentagonal NWs, we considered the [110] oriented pentagonal SnTe NWs for different thicknesses. We report the electronic band structures including the SOC effects for two cases of the [110] oriented SnTe pentagonal NWs with Sn on the twin boundaries and Te on the twin boundaries. The band structures of the SnTe and PbTe pentagonal NWs without SOC included are instead shown in Appendix D in the Supp. Information, no major differences are observed between SnTe and PbTe. However, the two pentagonal NWs with different twin boundaries show different properties as observed also in [111] thin films\cite{Samadi23}. In Fig. \ref{Bands_110_SOC} we show the two cases with different twin boundaries, where all the band structures reveal metallic behavior with bands connecting the valence and conduction electrons. While the behavior as a function of the thickness does not change for the Te twin-boundary NWs, for the 5-ring NW with Sn twin-boundary we observe an additional band connecting the valence and conduction band that seems to close and invert the band gap at the $\Gamma$ point.
We have a reduction of the band gap at $\Gamma$ but not at the Z point as a strong indication that the two high-symmetry k-points are different from the topological point of view.
In Appendix E in the Supp. Information, we show that the core chain (CC) of atoms is responsible for the metallicity in the systems. Indeed, the band connecting the valence and conduction electrons disappears when the core chain of atoms is removed within DFT calculations.

\begin{figure}[t]
\centering	
\includegraphics[width=\columnwidth,angle=0]{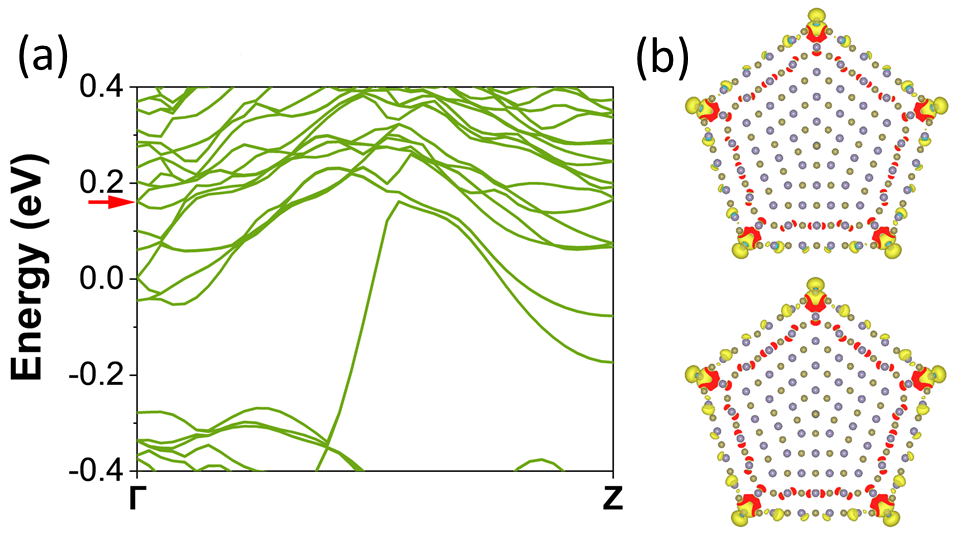}
\caption{(a) Band structure of the high-symmetric 5-ring SnTe pentagonal NW with Sn on the twin boundaries. The band connecting valence and conduction band has a positive slope.
The charge densities are shown (b) for the degenerate bands at the $\Gamma$ point. The red arrow in the panel (a) shows the two bands for which we extracted the charge density. The Fermi level is set to zero.}
\label{5ring_Sn}
\end{figure} 

\begin{figure}[t]
\centering	
\includegraphics[width=\columnwidth,angle=0]{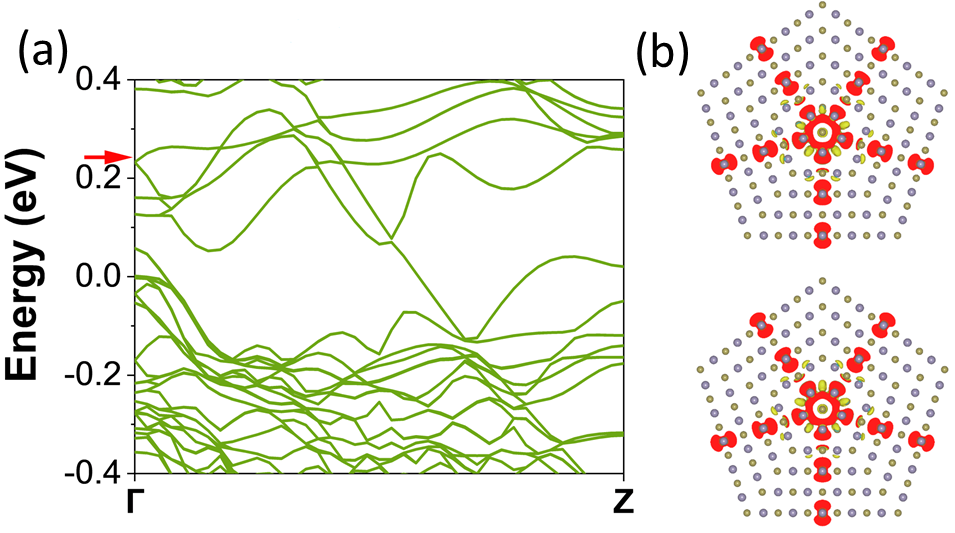}
\caption{(a) Band structure of the high-symmetric 5-ring SnTe pentagonal NW with Te on the twin boundaries.
The band connecting valence and conduction band has a negative slope. The charge densities are shown (b) for the degenerate bands at the $\Gamma$ point. The red arrow in the panel (a) shows the two bands for which we extracted the charge density. The Fermi level is set to zero.}
\label{5ring_Te}
\end{figure} 

In addition to the C$_5$ symmetry of the global system, it was shown that also the local C$_4$ symmetry is relevant.
Since the preservation of local C$_4$ symmetry in the region among the twinning is crucial for the observation of the higher-order topology\cite{10.21468/SciPostPhys.10.4.092}, we study the band structure of the pentagonal NWs with [110] orientation and perfect local C$_4$ symmetry (without structural relaxation). We define this crystal structure without structural relaxations as a high-symmetric pentagonal NW.
We show the electronic band structures and charge densities for the most symmetric 5-ring SnTe pentagonal NW with Sn and Te on the twin boundaries in Fig. \ref{5ring_Sn} and Fig. \ref{5ring_Te}. In the band structure of Fig. \ref{5ring_Sn}(a) and Fig. \ref{5ring_Te}(a) we observe a two-times degenerate band connecting the valence and conduction band with a positive slope for the Sn-twin boundary and negative slope for the Te-twin boundary.
In the case of the Te-twin boundary, we calculate the charge density at the $\Gamma$ point associated with the band connecting valence and conduction in Fig. \ref{5ring_Te}(b). Its charge is shared between the CC and the Sn atoms on the 5 mirror planes. Supported by the results presented in the model Hamiltonian, we assume that a similar state is present also in the Sn twin boundary NWs but in the conduction band and more hybridized with other bands.

\begin{figure}[t]
\centering	
\includegraphics[width=\columnwidth,angle=0]{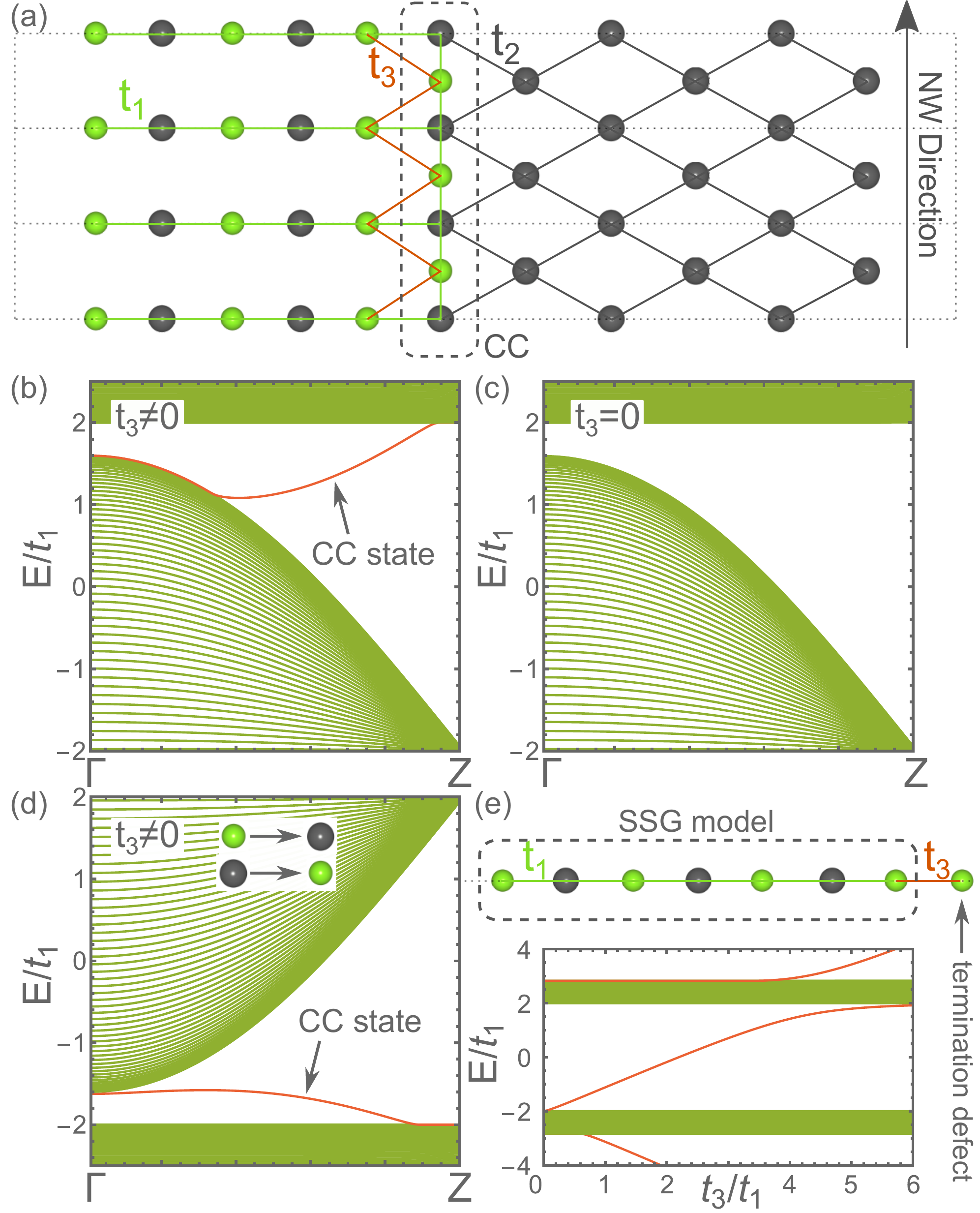}
\caption{Effective SSG-like ribbon model, the gray/green dots represent Sn and Te atoms with opposite on-site potentials $\pm$m defined in the model\cite{Brzezicki19}. (a) The transverse cross-section of the NW with the CC. Its length is equal to twice the number of rings. (b-c) The spectra of the ribbon system are periodic in the NW direction. In case (b) there is a non-vanishing hopping amplitude $t_3$ and we find a state strongly localized at the CC shown as a red line. In case of lack of this $t_3$ hopping (c) we do not find the band connecting valence and conduction band. (d) Same as (b) reversing anion and cation, in both (b) and (d) the Fermi level crosses the red line coming from the CC. (e) The minimal SSG model with termination defect and its spectrum showing end-states.
The parameters are: $m = 2t_1$, $t_2=t_3=0.9 t_1$.}
\label{SSG}
\end{figure}

To gain further insight, we perform tight-binding calculations using the model Hamiltonian for a section of the SnTe pentagonal NWs. The tight binding model was reported in previous literature\cite{Brzezicki19,Nguyen22} and it was derived from the Shiozaki-Sato-Gomi (SSG) model\cite{PhysRevB.91.155120} (for more see Appendix H in Supp. Information). We use as a section the mirror plane shown in Fig. \ref{SSG}(a) for a minimal model of the experimental system where the CC acts as a termination defect of the SSG model. 
In Fig. \ref{SSG}(b), the results for the Sn twin boundary describe a metallic band connecting the valence and conduction band associated with the CC. Even if this band is quite localized in the CC, the hybridization with the shell of the NWs described by the hopping t$_3$ is extremely relevant. Indeed, without the coupling between core and shell, the metallic band disappears as we can see in  Fig. \ref{SSG}(c).    
Exchanging anion and cation to change the twin boundary, we obtain a band structure for the Te twin-boundary shown in Fig. \ref{SSG}(d). With the Te twin boundary, we obtain the CC state to have a negative slope. Fig. \ref{SSG}(e) describes a simpler model with periodicity along the z-direction that further confirms the relevance of the CC as a termination defect of the SSG model for the formation of the metallic bands in pentagonal NWs.
Basically, in both twin boundary cases, the CC band connects the valence and conduction bands enforcing the metallicity even in the ultrathin limit. Therefore, the pentagonal NWs are intrinsic core-shell NWs composed of a single material with a metallic chain in the core and an insulating cubic SnTe shell. Core and shell are electronically and structurally different but have the same constituent elements and stoichiometry.

\subsection{Tunability of pentagonal nanowires growth}

Starting from the general aspects, we need to weaken the ionic bonds to favor the formation of pentagonal nanowires as we discussed in the previous subsections. Once the ionic part of the bond is weakened and the system is more covalent, the formation of pentagonal structures depends on the ratio of growth rates in the (100) and (111) directions, which is responsible for the twinning\cite{Wild1994,Buhler2000}. Many works on fivefold symmetry focus on nanoparticles. In addition to elemental materials, various covalent alloys, compounds, and composite nanoparticles grow with fivefold twinned structures\cite{Hofmeister2004}

Regarding the specific (Pb,Sn)Te system, we report what are the growth conditions that favor the formation of the pentagonal nanowires. We have hints of nucleation, but no conclusive evidence of nucleation of the central chain from several experiments (TEM of the NW//Si interface; STM at the center of the cross-section; detailed analysis of growth as a function of conditions). In the case of ternary (Pb,Sn)Te, the deviation of the chemical composition from x=0.5 (estimated for the layers based on the base of beam equivalent pressure) dramatically reduces the occurrence of 5-fold NWs. A change in the Te/Pb flux ratio of about 20\% from the optimal for 5-fold NWs growth results in growth NWs with 4-fold symmetry. Therefore, in order to observe the pentagonal nanowires the doping concentration should be around x=0.50 i.e. close to the critical concentration where there is the topological transition at room temperature\cite{Dornhaus1983}. Based on the experimental data, the slight excess of Pb in the total flux of (Pb,Sn)Te supports the nucleation of islands with other than a 4-fold structure, this agrees with our theoretical results supporting the presence of heavy atoms as necessary for the pentagonal NWs.
Regarding the temperature, for similar values of SnTe, Pb, and Te molecular fluxes, NWs 5-fold are observed at higher growth temperatures in the range where cubic NWs growth is not observed at all. Therefore, a high temperature is required since our theoretical results suggest that the pentagonal structure is metastable at low temperatures.

\begin{figure}[t]
\centering
\includegraphics[width=\columnwidth, angle=0]{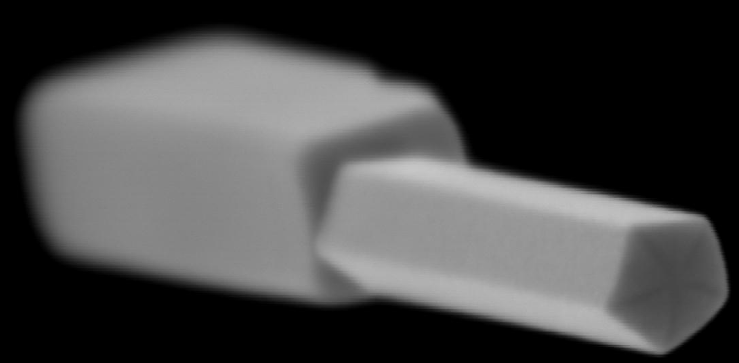}
	\caption{Heterostructure nanowire obtained from tuning the flux during the growth process to switch from the cubic to the pentagonal structure.}
	\label{cubic_pentagonal}
\end{figure}

Additionally, the crystal structure can be switched from cubic to pentagonal during the growth process to create heterostructure nanowires as shown in Fig. \ref{cubic_pentagonal}. In particular, this is the result of the PbTe/(Pb,Sn)Te//Si(100) growth process. In the first part of the growth, the growth time of (Pb,Sn)Te was performed, while in the second part, we grew PbTe. The growth of PbTe was carried out by turning off the SnTe flux for initial (Pb,Sn)Te and the PbTe flux remained unchanged. Hence, it suggests that both parts consist of (Pb,Sn)Te rather than PbTe/(Pb,Sn)Te. Further studies would be needed to establish the chemical composition of both parts of these NW heterostructures. More technical details are reported in Appendix A in the Supp. Information.

\begin{figure}[th]
\centering
\includegraphics[width=\columnwidth]{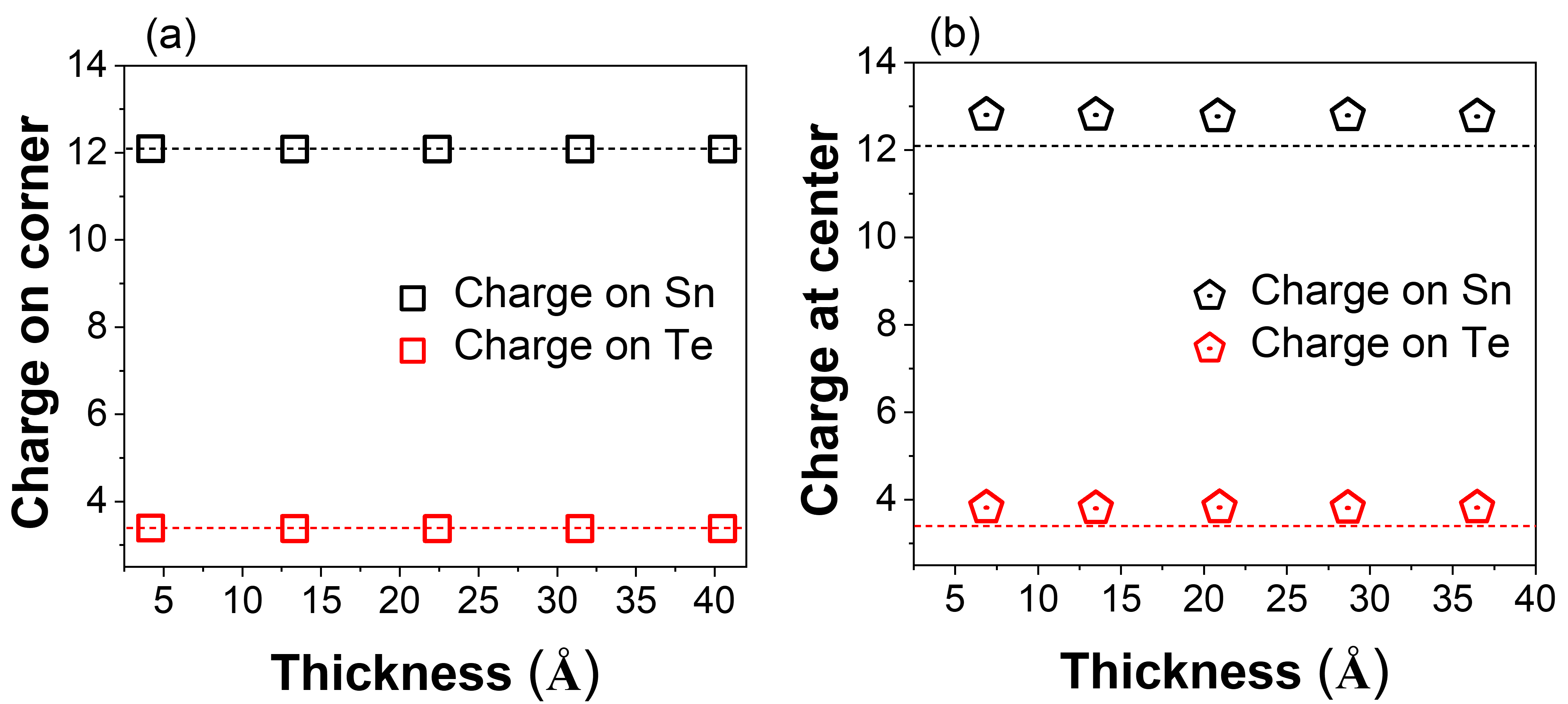}
\caption{Total charge on Sn and Te atoms as a function of thickness for the (a) trivial cubic NWs and (b) [110] pentagonal NWs. In the case of cubic systems, the charges are calculated for the corner atoms, whereas for pentagonal structures the charges for the core atoms are considered. The dashed horizontal lines represent the values of charges in the bulk SnTe, i.e. black for Sn and red for Te. 
}
\label{charge as a function of thickness}
\end{figure}

\subsection{Topological electronic states in the core chain of the pentagonal nanowires} 
 
Regarding the topological properties, we analyze two aspects of the pentagonal nanowires. In this subsection, we analyze the topological properties related to the core chain, while in the next subsection, we report the signature of the topological properties related to the fractional charge and hinge states.

The electronic properties of the core chain were understood using a simplified SSG model Hamiltonian derived from a ribbon of a topologically non-trivial model with nonsymmorphic chiral Z$_2$ invariant\cite{PhysRevB.101.235113}. This model (known also as the Rice-Mele model) was used also to study the presence of topological midgap corner modes in n-fold rotational symmetric chiral insulators to investigate the existence of a HOTI protected by rotational symmetry\cite{vanMiert2020}. The cross-section of the NW that was used for the model Hamiltonian is shown in Fig. \ref{SSG}(a), with the three hopping parameters of the model named as t$_1$, t$_2$ and t$_3$. When the hopping t$_3$ is zero, there are no bands connecting the valence and conduction band and the system is an insulator as shown in Fig. \ref{SSG}(c). When t$_3$ is different from zero, this model gives electronic states inside the gap connecting the valence and conduction bands resulting in a symmetry-enforced metallic phase. In the case of Sn on the twin boundary shown in Fig. \ref{SSG}(b), the connecting band has a positive slope while has a negative slope in the case of Te on the twin boundary plotted in Fig. \ref{SSG}(d). The robustness of this symmetry-enforced metallic phase was confirmed also by linear scaling DFT calculations (Appendix F in the Supp. Information) and by DFT calculations of the NWs under isotropic pressure (Appendix G in the Supp. Information). 
These results demonstrate that the electronic properties critically depend on the composition of the CC. 
The minimal model that reproduces this metallic phase is a SSG model with a termination defect as shown in Fig. \ref{SSG}(e). Note that a pure SSG model without termination defects exhibits no end-states 
\cite{PhysRevB.101.235113}. This is because the termination of the system breaks the nonsymmorphic chiral symmetry that protects the Z$_2$ topological invariant. However, in the presence of the topological invariant, there is charge accumulation at the edges, in strict analogy to the 1D systems with inversion symmetry\cite{PhysRevB.95.035421}. We find that with termination defect such accumulation can again lead to the appearance of the end-states.


\subsection{Signature of fractional charge and higher-order topology in pentagonal nanowires}

Now, we move to describe the signatures of higher-order topology as the hinge states. The hinge states for SnTe cubic nanowires (grown along the [001] orientation presenting a square section) were observed at a given k-point that was close to the edge of the Brillouin zone\cite{Nguyen22,Skiff22}.
In the case of the Sn twin boundary, we analyzed the charge density at the $\Gamma$ point in \ref{5ring_Sn}(b). Localized charge densities are visible at the corners of the 5-ring pentagonal NW. We surmise that this state will evolve in the localized higher-order topological state once the thickness of the shell is large enough to generate the TCI state of SnTe. While there exist recent studies focused on higher-order topology in finite systems in the xy plane, additional studies are necessary for the higher-order topology in NWs that are periodic along the z-direction and finite in the xy plane.

In addition to hinge states, another feature of the higher-order topology is the presence of a fractional charge at the center. Fig. \ref{charge as a function of thickness}(a) and \ref{charge as a function of thickness}(b) show the charge for the SnTe cubic and pentagonal NWs, respectively. We report for comparison also the charge in the bulk SnTe. The charges on both Sn and Te of the CC substantially increase with respect to the charge in the bulk and are constant as a function of the thickness. This excess of charge could be associated with the fractional charge appearing in the presence of disclinations. 
The fractional charge can be observed with and without topology. Fractional charge and higher-order topology are strongly developed in 2D materials but there are less evident in NWs. We leave it to future studies to provide for a deeper explanation of the fractional charge. The interplay between the topological properties from the core and from the shell will be described elsewhere.\cite{Samadi_Buczko}

\section{Conclusions}    

We have synthesized ionic pentagonal NWs of Pb$_{1-x}$Sn$_{x}$Te with smooth texture devoid of imperfections.
Within a semiclassical approach, the pentagonal crystal structure is stabilized by an interplay between the ionic bond and the covalent bond.
Within a full quantum approach, we have demonstrated that the pentagonal structure of the SnTe NWs is a metastable phase at T=0 but close in energy to the ground state which is the cubic structure. 
Therefore, we need to weaken the ionic bond with the presence of heavy atoms Pb and Te and to be close to the topological transition, finally, the growth of the pentagonal NWs should take place at high temperatures since they are not stable at low temperatures.
The pentagonal NWs are an intrinsic core-shell system where the core and shell have different structural, electronic and topological properties.
The pentagonal NWs with [110] orientations are metallic due to the presence of a single band connecting the conduction and valence bands coming from the one-dimensional disclination in the core of the NW. This connecting band clearly comes from the one-dimensional disclination and it is related to the properties of the topological SSG model. 
We have found an excess charge in the center that could be associated with a fractional charge. We have calculated the charge in the real space related to the degenerate electronic states at the $\Gamma$ point and we find that these states are associated with higher-order topological states at the hinges of the NWs.
The pentagonal NWs are a platform for higher-order topology and fractional charge. These pentagonal NWs represent
a unique case of intrinsic core-shell NWs with distinct structural, electronic and topological properties between the core and the shell.

\section*{Author contributions}
G. Hussain: Investigation, Methodology, Validation, Data curation, Writing - original draft. G. Cuono: Investigation, Methodology, Validation, Data curation, 
Writing - original draft. P. Dziawa: Investigation, Validation, Data curation, 
Writing - original draft.  D. Janaszko: Investigation, Validation, Data curation, 
Writing - original draft.  J. Sadowski: Funding acquisition. S. Kret: Funding acquisition. B. Kurowska: Investigation, Methodology, Validation,
J. Polaczyński: Conceptualization, Investigation, Methodology, Validation,
K. Warda: Investigation, Methodology, Validation, Data curation, 
Writing - original draft. S. Sattar: Writing - review \& editing. 
C. M. Canali: Funding acquisition, Resources, Writing - review \& editing. 
A. Lau: Validation W. Brzezicki: Methodology, Investigation, Validation, Data curation, Funding acquisition. T. Story: Supervision, Resources. Carmine Autieri: Supervision, Conceptualization, Project administration, Resources, Writing - review \& editing. 

\section*{Conflicts of interest}

The authors declare that they have no known competing financial interests or personal relationships that could have appeared to influence the work reported in this paper.

\section*{Acknowledgments}
We thank T. Dietl, C. Ortix, R. Buczko and S. Samadi for useful discussions.
The work is supported by the Foundation for Polish Science through the International Research Agendas program co-financed by the European Union within the Smart Growth Operational Programme (Grant No. MAB/2017/1). 
W.B. acknowledges support from Narodowe Centrum
Nauki (NCN, National Science Centre, Poland) Project No. 2019/34/E/ST3/00404. 
J.S. and S.K. acknowledge support from the National Science Centre Poland through the projects 2019/35/B/ST3/03381 and 2019/35/B/ST5/03434. C.M.C. and S.S. thank Carl Tryggers Stiftelsen (CTS 20:71) and the Swedish Research Council (VR 2021-04622) for financial support. 
We acknowledge the access to the computing facilities of the Interdisciplinary Center of Modeling at the University of Warsaw, Grant g91-1418, g91-1419 and g91-1426 for the availability of high-performance computing resources and support.
We acknowledge the CINECA award under the ISCRA initiative  IsC99 "SILENTS”, IsC105 "SILENTSG", IsB26 "SHINY" and IsB27 "SLAM"  grants for the availability of high-performance computing resources and support. We acknowledge the access to the computing facilities of the Poznan Supercomputing and Networking Center Grant No. 609 (pl0223-01) and pl0267-01. The computations at Linnaeus University were enabled by resources provided by the National Academic Infrastructure for Supercomputing
in Sweden (NAISS) at Dardel partially funded by the Swedish Research Council through grant agreement no. 2022-06725.\\

\scriptsize{
\bibliography{bibliography} 
\bibliographystyle{rsc} } 

\end{document}